\documentclass[preprint]{revtex4}
\usepackage{graphicx}
\begin{document}

\title{The origin of order in random matrices with symmetries}
%\classification{21.10.Hw,21.60.Cs,24.60.Lz}

%\keywords{random matrices, point-group symmetries}

\author{Calvin W. Johnson}
%{address={Department of Physics, San Diego State University,
%5500 Campanile Drive, San Diego, CA 92182-1233}}
\affiliation{Department of Physics, San Diego State University,
5500 Campanile Drive, San Diego, CA 92182-1233}

\begin{abstract}
From Noether's theorem we know  symmetries lead to conservation laws. What is left  to nature is the ordering of conserved quantities; for example, the quantum numbers of the ground state. In physical systems the ground state is generally associated with `low' quantum numbers and symmetric, low-dimensional irreps, but there is no \textit{a priori} reason to expect this.  By constructing random matrices with nontrivial point-group symmetries, I find the ground state is always dominated by extremal low-dimensional irreps. Going further, I suggest this explains the   dominance of $J=0$ g.s. even for random two-body interactions. 
\end{abstract}

\maketitle

One of the most `beautiful' results in mathematical physics is Noether's theorem. 
In classical mechanics a (continuous) symmetry leads to a conserved quantity, for 
example translational invariance leads to conservation of momentum, invariance 
under displacement in time leads to conservation of energy, and  
rotational invariance leads to conservation of angular momentum.  

In quantum mechanics we can think about symmetries in terms of groups and 
their irreps. Group irreps  divide up a Hilbert space into 
subspaces; if a Hamiltonian is invariant under a symmetry, meaning it commutes 
with the generators of a group, then the Hamiltonian becomes block-diagonal in 
the irreps. 

For continuous symmetries, we label the irreps by \textit{quantum numbers}, 
which in turn arise from the eigenvalues of the Casimir(s) of the symmetries.  
For discrete symmetries the Hamiltonian is still block-diagonal 
in the irreps, although lacking a Casimir there may not be a `natural' quantum number 
labeling the irreps. 

While Noether's theorem is a powerful result, it tells us nothing about the relative ordering 
of states.  In natural systems, however, we observe an ordering so striking and ubiquitous 
we tend to take it for granted, namely that the ground state and states lying low in the 
spectrum tend to belong to `small' quantum numbers with the most symmetric 
wavefunctions.  For example, under translational symmetry the ground state has 
momentum $p = 0$,  under rotational symmetry the ground state has $J =0$, etc.

Of course this arises because of physics: in most Hamiltonians the kinetic energy 
terms are quadratic in linear momentum $p$ and/or angular momentum $J$. This in turn occurs because Nature, or physicists, 
prefer \textit{almost}-local theories, and the first nontrivial non-local terms are quadratic in the gradient. 

But surprisingly, even when one removes all trace of `physics' the pattern remains. In 
 nuclear structure  this is seen in the discovery that many-body systems with 
rotationally invariant but 
otherwise random two-body Hamiltonians  nonetheless 
tend to have ground states with $J=0$, 
just like `realistic' interactions, even though such states are a small fraction 
of the total space\cite{JBD98,ZV04,ZAY04}.   
This phenomenon is robust; for example while textbooks traditionally 
ascribe the $J=0$ g.s. to the pairing interaction \cite{BM,simple}, $J=0$ still 
dominates the ground state even when the pairing matrix elements are all set to zero 
\cite{Jo99}.  It also occurs in boson systems \cite{BF00} such as the Interaction Boson 
Model\cite{IBM}. 
 Over the past decade there have been many  proposed `explanations.' 
As the distribution of many-body systems with two-body interactions tend to have a 
Gaussian distribution of states \cite{MF75}, a number of authors have focused on 
widths \cite{BFP99,PW04}, while others have statistically averaged in a single $j$-shell 
the coupling of
multiple angular momenta \cite{MVZ00}. As a 
recent Letter stated, `the simple question of symmetry and chaos asks for a simple 
answer which is still missing \cite{Vo08}.'

To investigate this phenomenon, I propose a novel approach.  In physics (as in art, 
that arbiter of `beauty') we often delve deeper by stripping away assumptions to 
see what remains.  
Previous studies of random two-body interactions used shell-model diagonalization 
 codes, but instead of carefully calculated matrix 
elements of two-body interactions they used random numbers,  insisting only
on rotational invariance.  Now I go a step further: I abandon the shell-model framework, 
take  random matrices, often used to investigate statistical properties of complex systems \cite{BM,RM},
and impose discrete rotational symmetries of the regular polyhedra upon them.

\begin{figure}  
\includegraphics [width = 7.5cm]{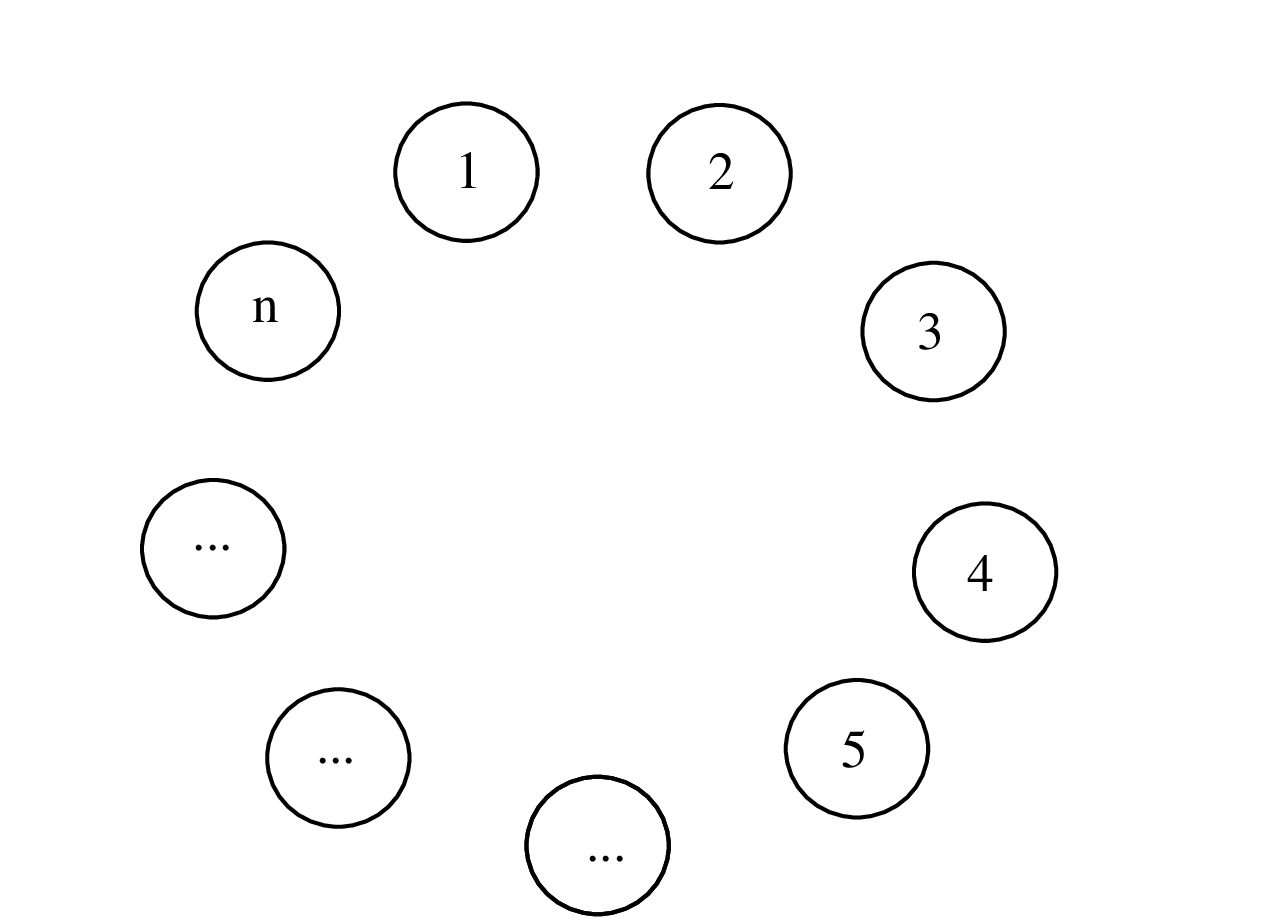}
\caption{\label{cn} 
Illustration of discrete rotation symmetry in the plane, $C_n$. }
\end{figure}

Let's start with $C_n$ symmetry, the symmetry of discrete rotations in a plane. Figure \ref{cn} illustrates with $n$ loci evenly spaced in a circle. The generator of 
discrete rotations is 
\begin{equation}
\mathbf{T} = 
\left (
\begin{array}{ccccccc}
0 & 0 & 0 & 0 & \ldots & 0  & 1 \\
1 & 0 & 0 & 0 & \ldots & 0  & 0 \\
0 & 1 & 0 & 0 & \ldots & 0 &  0  \\
\vdots & & & & & & \\
0 & 0 & 0 & 0 & \ldots & 1 & 0 
\end{array}
\right ) \label{Tdef}
\end{equation}
which send $1 \rightarrow 2$, $2 \rightarrow 3$ and so on. One can easily show by 
hand that the most general real Hamiltonian invariant under discrete rotations 
$\mathbf{H} = \mathbf{T} \mathbf{H} \mathbf{T}^{-1}$ is
\begin{equation}
\mathbf{H} = 
\left (
\begin{array}{ccccccc}
a & b & c & d & \ldots & c & b \\
b & a & b & c & \ldots & d  & c \\
c & b & a & b & \ldots & e  & d  \\
\vdots & & & & & & \\
b & c & d & e & \ldots & b  & a 
\end{array}
\right ) \label{hdef1}
\end{equation}
where $a,b,c, \ldots$ are random numbers.  (Note: here and throughout I do not 
consider adding reflection symmetries, saving that for future work.)

One can solve this exactly without diagonalization by  
discrete Fourier transforms. 
One  writes the matrix element $H_{i,j}$ in terms of a real 
function of a single index,
$H_{i,j} = F_{i-j},$
so that, in (\ref{hdef1}), $F_0 = a, F_1 = b, F_2 = c$, etc..
The function $F_{-j} = F_j$ from hermiticity, and by further
inspection one gleans $F_{n-j} = F_{j+1}$.

The next step is the Fourier decomposition of $F_j$:
\begin{equation}
F_j = \frac{1}{n} \sum_{k = 0}^{n-1} h_k \exp(i2\pi k j/n).
\end{equation}
Inverting,
\begin{equation}
h_k = \sum_{j = 0}^{N-1} F_j \exp ( -i2\pi kj/n)
\end{equation}
but using $F_{n-j} = F_{j+1}$
\begin{equation}
h_k = 
F_0 + \sum_{j = 1}^{n-1} F_n  \cos (2 \pi k j/n) =
F_0 + \sum_{j = 1}^{[n/2] } F_n \zeta_{j,n} \cos (2 \pi k j/n)
\label{cneigen}
\end{equation}
where
$[x]$ is the floor function and $\zeta_{j,n} = (2-\delta_{j,n/2+1})$ prevents double-counting
when $n$ is even. It is easy to verify that $h_k = h_{n-k}$.
The $h_k$ are the eigenvalues of $\mathbf{H}$, 
 and one is justified in using $k$ to label the different irreps.  One can also find 
the eigenvectors $\vec{\psi}^{(k)}$ easily;  $k=0$ is the most symmetric,
with ${\psi}^{(k)}_j \propto $ constant, while $k=n/2$ is arguably the least symmetric, with 
${\psi}^{(k)}_j \propto (-1)^j$.

While (\ref{cneigen}) gives the eigenvalues from a simple sum, we consequently have no 
\textit{a priori} way of identifying the 
ground state energy which irrep it belongs to. 

Next I assume there are additional degrees of 
freedom, as yet unspecified, and replace the scalars $a, b, c, \ldots$ by 
$m \times m$ real symmetric matrices $\mathbf{A}, \mathbf{B}, \mathbf{C}, \ldots$ so that 
\begin{equation}
\mathbf{H} = 
\left (
\begin{array}{ccccccc}
\mathbf{A} & \mathbf{B} & \mathbf{C} & \mathbf{D} & \ldots & \mathbf{C} & \mathbf{B} \\
\mathbf{B} & \mathbf{A} & \mathbf{B} & \mathbf{C} & \ldots & \mathbf{D} & \mathbf{C} \\
\mathbf{C} & \mathbf{B} & \mathbf{A} & \mathbf{B} & \ldots & \mathbf{E}  & \mathbf{D}  \\
\vdots & & & & & & \\
\mathbf{B} & \mathbf{C} & \mathbf{D} & \mathbf{E} & \ldots & \mathbf{B}  & \mathbf{A} 
\end{array}
\right ) \label{hdef2}
\end{equation}
a matrix of dimension $mn \times mn$.

This matrix can no longer be immediately solved. What we \textit{can} do, however, 
is to put $\mathbf{H}$ into block-diagonal form, that is, 
\begin{equation}
\mathbf{H} = 
\left (
\begin{array}{ccccccc}
\mathbf{h}_0 & 0 & 0 & 0 & \ldots & 0 & 0 \\
0 & \mathbf{h}_1 & 0 & 0 & \ldots & 0  & 0 \\
0 & 0 & \mathbf{h}_2 & 0 & \ldots & 0  & 0  \\
\vdots & & & & & & \\
0 & 0 & 0 & 0 & \ldots & 0  & \mathbf{h}_n 
\end{array}
\right )
\end{equation}
where
\begin{equation}
\mathbf{h}_k =\mathbf{F}_{0}+ \sum_{n = 1}^{[n/2]}
\mathbf{F}_{j} \zeta_{j,n} \cos ( 2 \pi kj/n)
\end{equation}

Here comes the key step.   While we cannot analytically compute the ground state 
energy of each $\mathbf{h}_k$, we can  compute the variance.  If we assume the 
$\mathbf{F}_j$ are independent random matrices, each with the same variance $\sigma_0^2$ (assumptions 
which can be relaxed), then the variance of the $k$th block is 
\begin{equation}
\sigma^2_k = \sigma^2_{0}\left(1+ \sum_{j = 2}^{[n/2]+1} \zeta_{j,n}^2 \cos^2 ( 2 \pi k(1-j)/n)\right)
\end{equation}
For $k = 0$, this yields approximately
$\sigma^2_{0} (2n+1)$, while otherwise this will yield approximately $\sigma^2_{0}
(n+1)$ (because the average of $\cos^2 x \approx 1/2$). Thus the matrices with $k=0$ will have the
larger widths and the ground state $k$ will be one of those two.

\begin{figure}  
\includegraphics [width = 7.5cm]{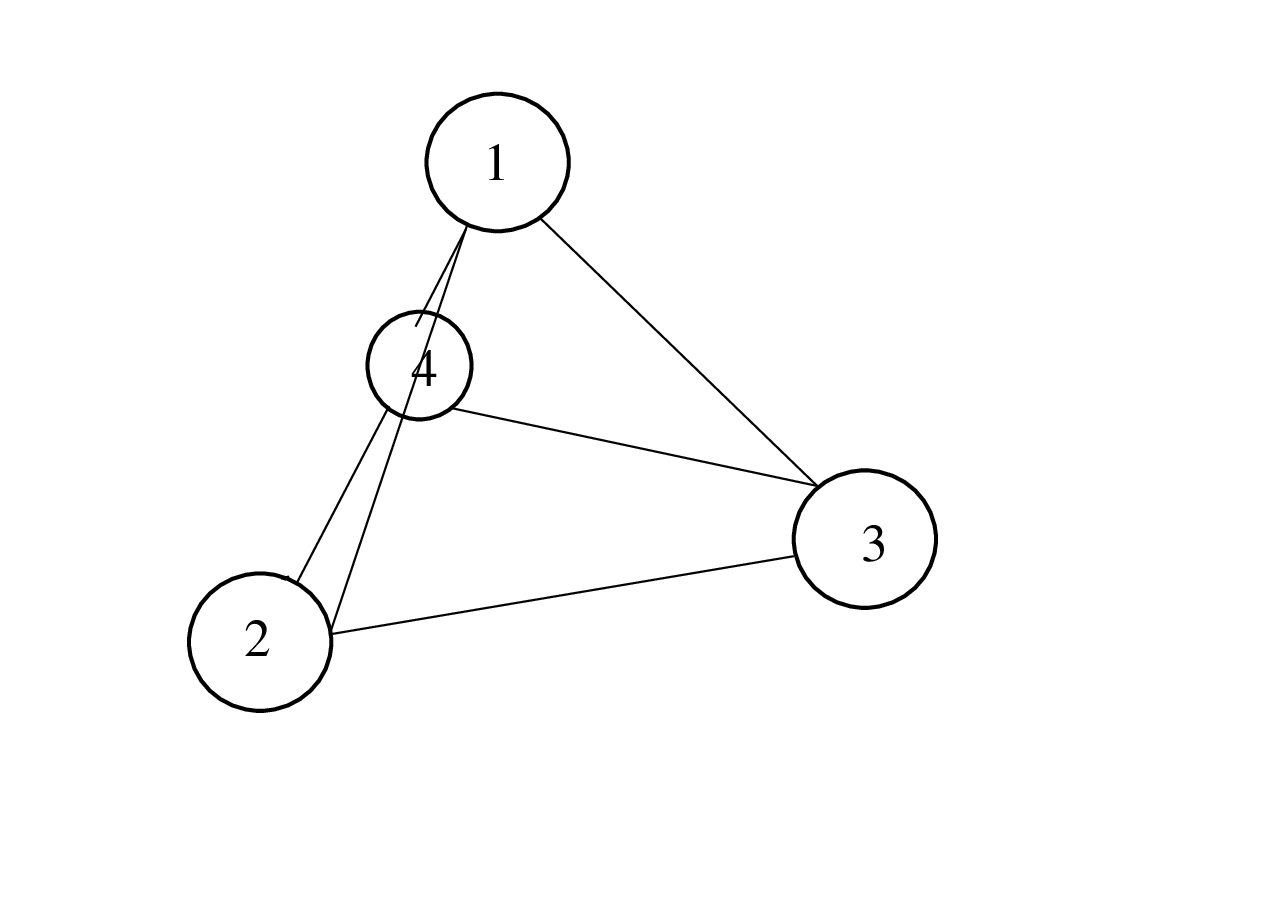}
\caption{\label{tet} 
Illustration of tetrahedron space symmetry. }
\end{figure}

Now $C_n$ is an abelian group, but we can do the same analysis for other,
nonabelian point-symmetry groups based on regular polyhedra. 
First, consider the tetrahedron, illustrated in Fig.~\ref{tet}.  The most general Hamiltonian invariant 
under any discrete rotation about any of its facets is
\begin{equation}
\left(
\begin{array}{cccc}
\mathbf{A} &\mathbf{B}  &\mathbf{B} &\mathbf{B} \\
\mathbf{B}  & \mathbf{A} &\mathbf{B} &\mathbf{B} \\
\mathbf{B}  &\mathbf{B} &\mathbf{A} &\mathbf{B} \\
\mathbf{B}  &\mathbf{B} &\mathbf{B} & \mathbf{A} 
\end{array} \right )
\end{equation}
Bringing this into block diagonal form, 
\begin{equation}
\left(
\begin{array}{cccc}
\mathbf{A}+ 3\mathbf{B} &0 &0 &0 \\
0  & \mathbf{A} - \mathbf{B} & 0 & 0  \\
0  & 0 &\mathbf{A}  - \mathbf{B} & 0  \\
0 & 0  &0  & \mathbf{A}  - \mathbf{B}
\end{array} \right )
\end{equation}
That is, there is an irrep of  dimension 1 with matrix $\mathbf{A}+ 3\mathbf{B}$ 
and an irrep of dimension 3 with matrix $\mathbf{A}-\mathbf{B}$.
Again assuming $\mathbf{A}, \mathbf{B}$ are independent but have the same variance 
the variance of the 1-dimensional irrep (which has the most symmetric 
eigenvector)  is $10 \sigma^2_0$ while that of the 3-dimensional 
irrep is $2 \sigma^2_0$.  All else being equal, the ground state is much more likely belong to the 
1-dimensional irrep.

\begin{figure}  
\includegraphics [width = 7.5cm]{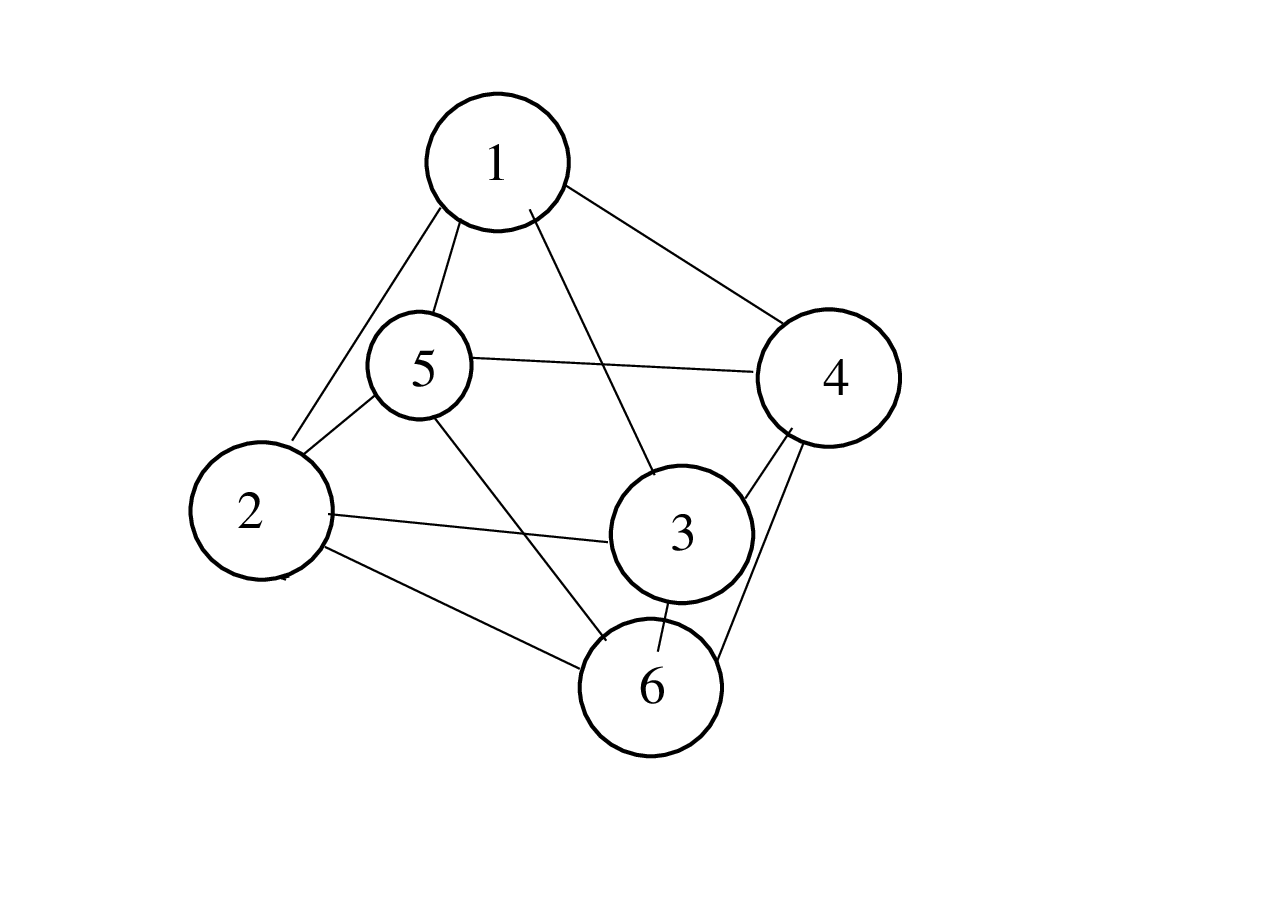}
\caption{\label{oct} 
Illustration of octahedral space symmetry. }
\end{figure}
With the basic idea in hand, it is easy to consider other polyhedra. 
For the octahedron, Fig.~\ref{oct}, the invariant hamiltonian is 
\begin{equation}
\left(
\begin{array}{cccccc}
\mathbf{A} &\mathbf{B}  &\mathbf{C}  &\mathbf{B}  & \mathbf{B} &\mathbf{B} \\
\mathbf{B}  & \mathbf{A} &\mathbf{B} &\mathbf{C}  & \mathbf{B} &\mathbf{B} \\
\mathbf{C}  &\mathbf{B} &\mathbf{A} &\mathbf{B}  & \mathbf{B}  &\mathbf{B} \\
\mathbf{B}  & \mathbf{C}  &\mathbf{B}  & \mathbf{A} & \mathbf{B} & \mathbf{B} \\ 
\mathbf{B}  & \mathbf{B}  &\mathbf{B}  & \mathbf{B} & \mathbf{A} & \mathbf{C} \\ 
\mathbf{B}  & \mathbf{B}  &\mathbf{B}  & \mathbf{B} & \mathbf{C} & \mathbf{A}
\end{array} \right )
\end{equation}
which has a 1-dimensional irrep, with matrix $\mathbf{A}+ 4\mathbf{B} + \mathbf{C}$, 
a 2-dimensional irrep with matrix $\mathbf{A}-2\mathbf{B} + \mathbf{C}$, and 
a 3-dimensional irrep with matrix $\mathbf{A}- \mathbf{C}$. The variances are 
$18\sigma^2_0$, $6\sigma^2_0$, and $2\sigma^2_0$, respectively.

\begin{figure}  
\includegraphics [width = 7.5cm]{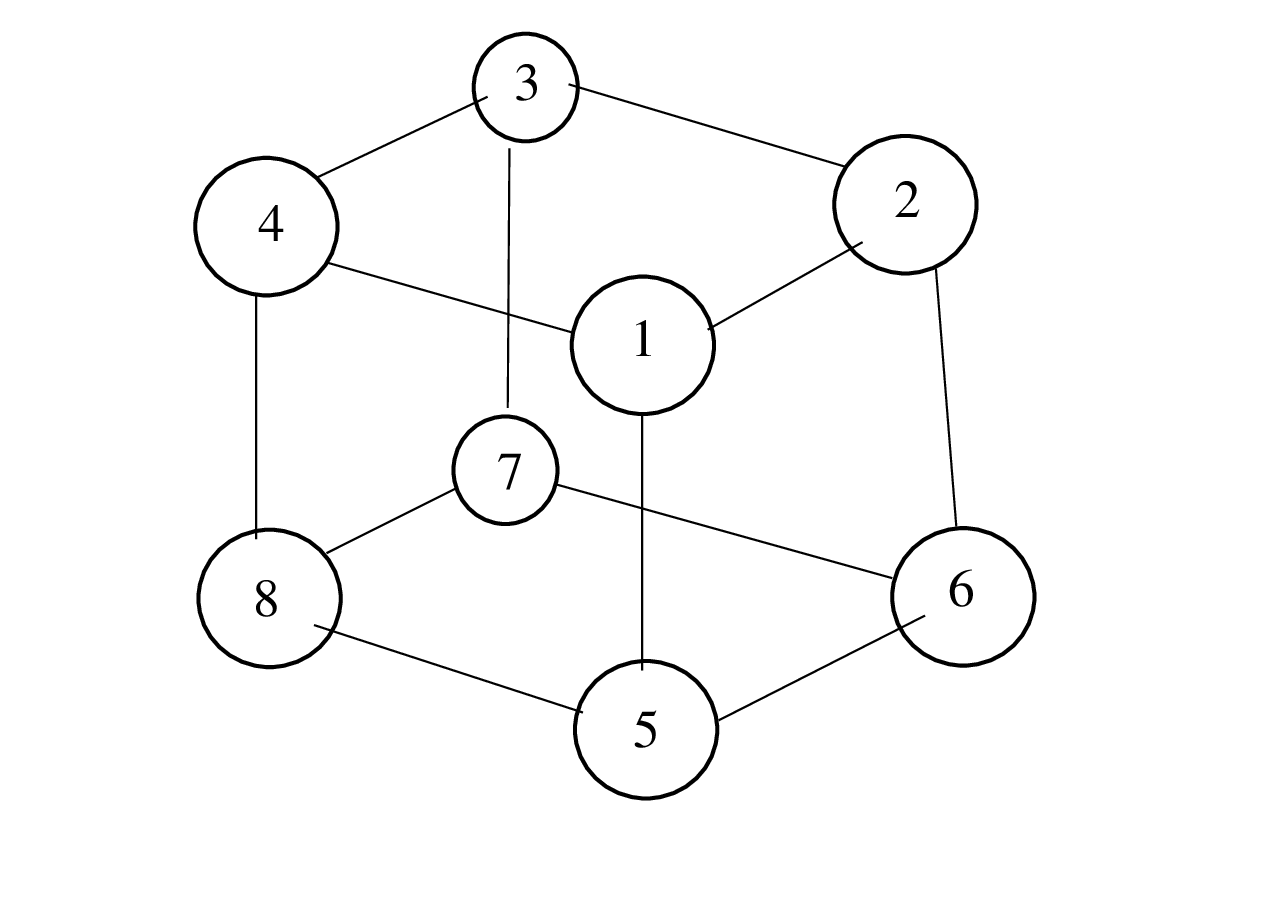}
\caption{\label{cube} 
Illustration of cubic space symmetry. }
\end{figure}

For the cube, Fig. \ref{cube}, the Hamiltonian is of the form
\begin{equation}
\left(
\begin{array}{cccccccc}
\mathbf{A} &\mathbf{B}  &\mathbf{C}  &\mathbf{B}  & \mathbf{B} &\mathbf{C } &  \mathbf{D} & \mathbf{C} \\
\mathbf{B} &\mathbf{A}  &\mathbf{B}  &\mathbf{C}  & \mathbf{C} &\mathbf{B } &  \mathbf{C} & \mathbf{D} \\
\mathbf{ C } &\mathbf{B  }  &\mathbf{A  }  &\mathbf{B }  & \mathbf{ D} &\mathbf{C} & \mathbf{B} & \mathbf{C} \\
\mathbf{B} &\mathbf{C }  &\mathbf{B }  &\mathbf{A }  & \mathbf{ C} &\mathbf{D} & \mathbf{C} & \mathbf{B } \\
\mathbf{B} &\mathbf{C } &\mathbf{D } &\mathbf{C } & \mathbf{A } &\mathbf{B } &  \mathbf{C } & \mathbf{B } \\
\mathbf{ C } &\mathbf{ B }  &\mathbf{C  }  &\mathbf{ D  }  & \mathbf{ B } &\mathbf{B  } &  \mathbf{B  } & \mathbf{C  } \\
\mathbf{D  } &\mathbf{C  }  &\mathbf{B}  &\mathbf{ C}  & \mathbf{ C } &\mathbf{B} &  \mathbf{A} & \mathbf{B} \\
\mathbf{ C} &\mathbf{ D}  &\mathbf{C  }  &\mathbf{ B }  & \mathbf{B} &\mathbf{C  } &  \mathbf{B  } & \mathbf{ A } 
\end{array} \right )
\end{equation}
with two 1-dimensional irreps with matrices $\mathbf{A} \pm 3 \mathbf{B} + 3 \mathbf{C} \pm \mathbf{D}$ 
and two 3-dimensional irreps with matrices $\mathbf{A}-\mathbf{C} \pm (\mathbf{B} - \mathbf{D} )$.  
The 1-dimensional irreps have variance $20 \sigma_0^2$ while the 3-dimensional irreps have 
variance $4\sigma_0^2$. 

While it remains to be proved in general, the lesson is clear: Starting with random 
matrices and imposing symmetries, the ground state is naturally dominated by 
certain irreps, generally irreps with lowest dimension. 

%%%%%%%%%%%%%%%%%%%%%%%%%%%% OLD STUFF %%%%%%%%%%%%%%%%%%%%%%%%%%%%%%%%%%%%%%

Now I turn to continuous symmetries such as SU(2).
I consider wavefunctions that can be written in the form 
$\psi_{l}(\vec{v}) Y_{lm}(\theta, \phi)$
where all of the rotational information is bound up in the spherical 
harmonic $Y_{lm}$ and $\vec{v}$ refers to internal degrees of freedom 
\cite{SUV98}.  I'll return to the latter in a moment. 
Using the angles
from spherical coordinates, the Hamiltonian is of 
the form ${H}(\theta^\prime \phi^\prime, 
\theta \phi)$, but imposing rotational invariance means $\hat{H}$ can only 
depend on the angle $\omega$ \textit{between} $\theta^\prime, \phi^\prime$ and
$\theta, \phi$ as given by 
$\cos \omega = \cos \theta \cos \theta^\prime + \sin \theta \sin \theta^\prime 
\cos( \phi - \phi^\prime)$. Then 
${H}(\theta^\prime \phi^\prime, \theta \phi) = F( \omega),$ where $F(\omega) = F(\omega+2\pi)$ is a periodic function and, using Hermiticity 
(and assuming $H$ is real) $F(\omega) = F(2\pi-\omega)$ 
is symmetric with respect to $\omega = \pi$. 
Expanding
\begin{eqnarray}
F(\omega) = \sum_J h_J\frac{2J+1}{4\pi} P_J(\cos \omega) \nonumber \\
= \sum_J h_J \sum_{M = -1}^J Y_{JM}(\theta^\prime,\phi^\prime) Y^*_{JM}(\theta,\phi)
\end{eqnarray}
so clearly the $h_J$ are again the eigenvalues, with $Y_{JM}$ as eigenfunctions 
and with the eigenvalues independent of $N$, as one expects.  
%Note: I use $J$ to denote generic total angular momentum, which may be composed from both orbital and spin angular momenta. 

Once more I assume $\mathbf{F}(\omega)$ to be a matrix-valued function, and 
thus $\mathbf{h}_J$ to be a symmetric matrix given by
\begin{equation}
\mathbf{h}_J = 2\pi \int_0^\pi P_J(\cos \omega) \mathbf{F}(\omega) \sin \omega d\omega.
\end{equation}
As before, let $\bar{\sigma}$ be the variance of the matrix elements $\mathbf{F}$ independent
of $\omega$. 
Then I estimate the variance of the matrix elements of $\mathbf{h}_J$ 
\begin{equation}
\sigma^2_J = 4\pi^2 \bar{\sigma}^2 \int_0^\pi P_J^2(\cos \omega) \sin^2 \omega d\omega 
\label{SU2variance}
\end{equation}
%where the weighting factor $\sin^2 \omega$ arises because when I discretize, I assume the fluctuations in $\mathbf{F}$ should be binned  with respect to $\omega$ and not $\cos \omega$.  
Eqn.~(\ref{SU2variance}) can be computed numerically, and leaving off the 
factor $4\pi^2 \bar{\sigma}^2$, the  values are 
1.571, 0.393,  0.245, 0.178, 0.139  for $J = 0,1,2,3,4$ respectively. 

This suggests that in many-body system, subspaces with low-valued quantum numbers
will have larger widths. 
But in realistic, finite many-body calculations, subspaces with 
different $J$s have different dimensions. 
Furthermore, in the above argument each
$\mathbf{h}_J$ has independent random matrix elements, 
which typically has a semi-circular density of states \cite{RM}, 
yet for many-body systems with only two-body interactions the 
density of states tends towards a Gaussian\cite{MF75}.

I can approximately correct both deficiencies. First, following standard 
results on matrices with Gaussian-distributed matrix elements \cite{RM}, I let
$\sigma_J^{\mathrm{eff}} = \sqrt{N_J} \sigma_J \label{scaledwidth}$ be the width of the subspace of states with angular momentum $J$. 
Then, for each $J$, I simply create $N_J$  energies via a random Gaussian distribution 
of width 
$\sigma_J^{\mathrm{eff}}$, and ultimately determine the fraction 
of ground states with angular momentum $J$. 

\begin{figure}  
\includegraphics [width = 7.5cm]{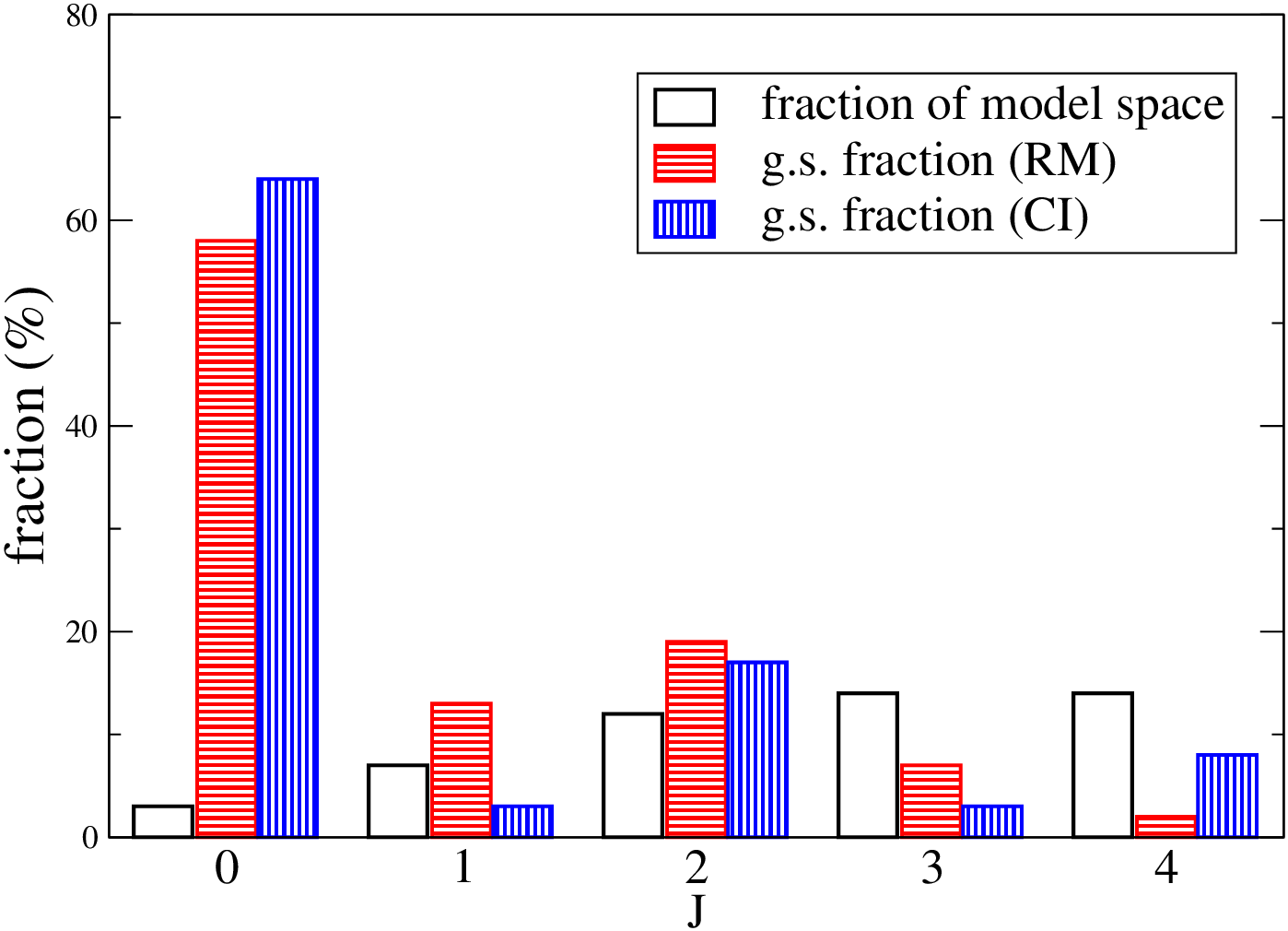}
\caption{\label{ca} (color online) 
Distribution of ground state quantum numbers
for eight neutrons in the $1p$-$0f$ shell. 
Empty bars are the fraction of states in the 
model space with a given $J$, horizontally striped (red) bars 
are the fraction of ground states with a given $J$ predicted 
by a random matrix (RM) model, and vertically striped (blue) bars 
are fraction of ground states with a given $J$ from 
configuration-interaction (CI) diagonalizations of an ensemble
of random two-body 
Hamiltonian in a shell-model basis. }
\end{figure}

Finally, I compare against a simulation via configuration-interaction
calculations, that is, diagonalizing 
the Hamiltonian for fixed numbers of 
particles in finite single-particle spaces. 
Figure \ref{ca} shows the case of eight fermions (neutrons) in the 
$1p_{1/2}$-$1p_{3/2}$-$0f_{5/2}$-$0f_{7/2}$ shell-model space.

 I take an 
ensemble of rotationally invariant, two-body but otherwise random interactions, 
and tabulate the fraction $f_\mathrm{CI}$ of states that have a given 
angular momentum $J$.  This should be compared with the native fraction of states with 
that $J$ in each many-body space, $f_\mathrm{space} = N_J/N_\mathrm{tot}$ 
($N_\mathrm{tot}$ is the total dimension of the many-body space), and  the 
fraction with $J=0$ is dramatically enhanced. 
I also compare with the fraction of ground states with a given $J$ predicted by my 
simple random matrix picture, $f_\mathrm{RM}$. The only input are the dimensions $N_J$
 and the universal variances computed in
(\ref{SU2variance}).

For such a simple picture, the random matrix model yields qualitatively 
excellent results, generally predicting the enhancement or suppression of different 
$J$s in the CI simulations relative to the native fractions $f_\mathrm{space}$. 
The RM model successfully predicts an enormous enhancement 
of $J=0$ in the ground state, for this case and many others not shown due to lack of space.

This analysis suggests the predominance of angular-momentum zero 
ground states is primarily a function of the width of the angular-momentum-projected 
many-body Hamiltonian; furthermore, the width is largely decoupled from the 
microphysics, instead depending only on the projection integrand  (\ref{SU2variance}) 
and on the dimensionality of subspaces with good quantum numbers. 
The simplicity and decoupling from the microphysics may be why the phenomenon is 
so robust and so universal.

%  CONCLUSION

%It will be 
%interesting to see what further insights  can be gleaned using this approach. 
%Further investigation would be particularly warranted into symmetry breaking, two-species 
%systems, and correlations between states with different quantum numbers. 

%\begin{theacknowledgments}

The U.S.~Department of Energy supported this investigation through
grant DE-FG02-96ER40985. 
%\end{theacknowledgments}

% I thank Erich Ormand for installing isospin-breaking capability into the BIGSTICK code. 

\bibliographystyle{aipprocl} % if natbib is missing

\end{document}